\documentclass[11pt,preprint]{aastex}

\newcommand{\ergsec}{\mbox{erg s$^{-1}$}}
\newcommand{\Lx}{\mbox{$L_{\rm X}$}}
\newcommand{\chandra}{\mbox{\textit{Chandra}}}
\newcommand{\rosat}{\mbox{\it ROSAT}}
\newcommand{\cxo}{\textit{Chandra X-ray Observatory}}

\newcommand{\xmm}{{\it XMM-Newton}}

\begin{document}

\title{Dynamical Formation of Close Binary Systems in Globular Clusters}
\slugcomment{Submitted to {\it The Astrophysical Journal}}
\shorttitle{Close Binaries in Globular Clusters}
\shortauthors{Pooley et al.}

\author
{David Pooley\altaffilmark{1},
Walter H.\ G.\ Lewin\altaffilmark{1},
Scott F.\ Anderson\altaffilmark{2},
Holger Baumgardt\altaffilmark{3},
Alexei V.\ Filippenko\altaffilmark{4},
Bryan M.\ Gaensler\altaffilmark{5},
Lee Homer\altaffilmark{2},
Piet Hut\altaffilmark{6},
Victoria M.\ Kaspi\altaffilmark{1,7},
Junichiro Makino\altaffilmark{3},
Bruce Margon\altaffilmark{8},
Steve McMillan\altaffilmark{9},
Simon Portegies Zwart\altaffilmark{10},
Michiel van der Klis\altaffilmark{10},
Frank Verbunt\altaffilmark{11}}

\altaffiltext{1}{Center for Space Research, Massachusetts Institute of Technology,
Cambridge, Massachusetts 02139; dave@mit.edu, lewin@space.mit.edu}
\altaffiltext{2}{Astronomy Department, University of
Washington, Seattle, Washington 98195; anderson@astro.washington.edu, homer@astro.washington.edu}
\altaffiltext{3}{Department of Astronomy, University of Tokyo, Tokyo 113-0033, Japan; holger@astron.s.u-tokyo.ac.jp, makino@astron.s.u-tokyo.ac.jp}
\altaffiltext{4}{Department of Astronomy, University of California, Berkeley,
California 94720; alex@astron.berkeley.edu}
\altaffiltext{5}{Harvard-Smithsonian Center for Astrophysics, Cambridge, Massachusetts
02138; bgaensler@cfa.harvard.edu}
\altaffiltext{6}{Institute for Advanced Study, Princeton, New Jersey 08540; piet@ias.edu}
\altaffiltext{7}{Physics Department, McGill University, Montreal, Quebec, Canada; vkaspi@hep.physics.mcgill.ca}
\altaffiltext{8}{Space Telescope Science Institute, Baltimore, Maryland 21218; margon@stsci.edu}
\altaffiltext{9}{Department of Physics, Drexel University, Philadelphia, Pennsylvania
19104; steve@kepler.physics.drexel.edu}
\altaffiltext{10}{Astronomical Institute ``Anton Pannekoek,'' University of Amsterdam,
1098 SH Amsterdam, The Netherlands; spz@science.uva.nl, michiel@astro.uva.nl}
\altaffiltext{11}{Astronomical Institute, Utrecht University, 3508 TA Utrecht, The
Netherlands; f.w.m.verbunt@astro.uu.nl}

\begin{abstract}
We know from observations that globular clusters are very efficient
catalysts in forming unusual short-period binary systems or their
offspring, such as low-mass X-ray binaries (LMXBs; neutron stars
accreting matter from low-mass stellar companions), cataclysmic
variables (CVs; white dwarfs accreting matter from stellar
companions), and millisecond pulsars (MSPs; rotating neutron stars
with spin periods of a few ms).  Although there has been little direct
evidence, the overabundance of these objects in globular clusters has
been attributed by numerous authors to the high densities in the
cores, which leads to an increase in the formation rate of exotic
binary systems through close stellar encounters.  Many such close
binary systems emit X-radiation at low luminosities ($\Lx \lesssim
10^{34}~\ergsec$) and are being found in large numbers through
observations with the \cxo.  Here we present conclusive observational
evidence for a link between the number of close binaries observed in
X-rays in a globular cluster and the stellar encounter rate of the
cluster.  We also make an estimate of the total number of LMXBs in
globular clusters in our Galaxy.
\end{abstract}


\maketitle

Since the first evidence from the {\it Uhuru} and OSO-7 satellites
revealed a population of highly luminous ($\Lx \gtrsim
10^{36}~\ergsec$) LMXBs in globular clusters, it has been noted that
the formation rate per unit mass of these objects is orders of
magnitude higher in globular clusters than in the Galactic disk
\citep{ka75,cl75}.  This discovery stimulated a flurry of theoretical
work into the formation of globular cluster LMXBs by the processes of
two- and three-body encounters
\citep{ka75,cl75,fpr75,su75,hi75,he75,hi76}.  These dynamical
formation scenarios (as opposed to the independent evolution of
primordial binaries) are a natural explanation for the high occurrence
of LMXBs in globular clusters since the stellar densities, and hence
encounter rates, are much higher in the cores of globulars than other
regions of the Galaxy.  \citet{vh87} showed that the 11 bright LMXBs
known at that time in globular clusters (currently, there are 13
known; \citealt{wh01}) were consistent with being formed dynamically
through close encounters.

The population of close binaries in a globular cluster, in turn,
exerts a great influence on the dynamical evolution of the cluster.
Heggie's law \citep{he75} tells us that close binaries tend to become
even closer, on average, through encounters with single stars or other
less close binaries.  While doing so, they increase their binding
energy by transferring significant energy to other stars in their
environment.  Even a modest population of primordial binaries contains
a potential reservoir of binding energy that easily exceeds the
kinetic energy of all single stars in the cluster.  One of the
consequences is that primordial binaries can postpone deep core
collapse \citep{gh89}.  For a general introduction and review, see
\citet{hh03}.

The interplay between stellar dynamics and stellar evolution, as
external and internal factors modifying the binary properties, is
highly complex, and many details of these processes are not well
understood \citep{hu03,si03}.  The aim of this publication is to
employ recent X-ray data to focus in a reasonably model-independent
way on the gross environmental effects that clusters exert on their
binary population, and in turn on the feedback of the binaries in
changing their environments.  Our approach is to study the close
binary populations of clusters which differ greatly in their physical
properties.  This has only recently become feasible, due in large part
to the \cxo.

As the \chandra\ observations of 47~Tuc \citep{gr01a},
$\omega$~Centauri \citep{rut02}, NGC~6397 \citep{gr01b}, NGC~6440
\citep{po02b}, NGC~6626 \citep{be03}, and NGC~6752 \citep{po02a} have
shown, high spatial resolution X-ray images are one of the most
effective methods of finding large numbers of close binaries in
globular clusters since many of them (quiescent LMXBs, CVs, MSPs, and
coronally active main-sequence binaries) are low-luminosity X-ray
emitters.  This population of low X-ray luminosity sources was first
discovered by \citet{hg83a,hg83b} using data from the first fully
imaging X-ray satellite, the {\it Einstein Observatory}.  Later
observations with the X-ray imaging satellite \rosat\ expanded the
known population of these objects; in observations of 55 globular
clusters with \rosat, 57 low X-ray luminosity sources were discovered
\citep{vb00}.

\begin{figure}
\centering
\plotone{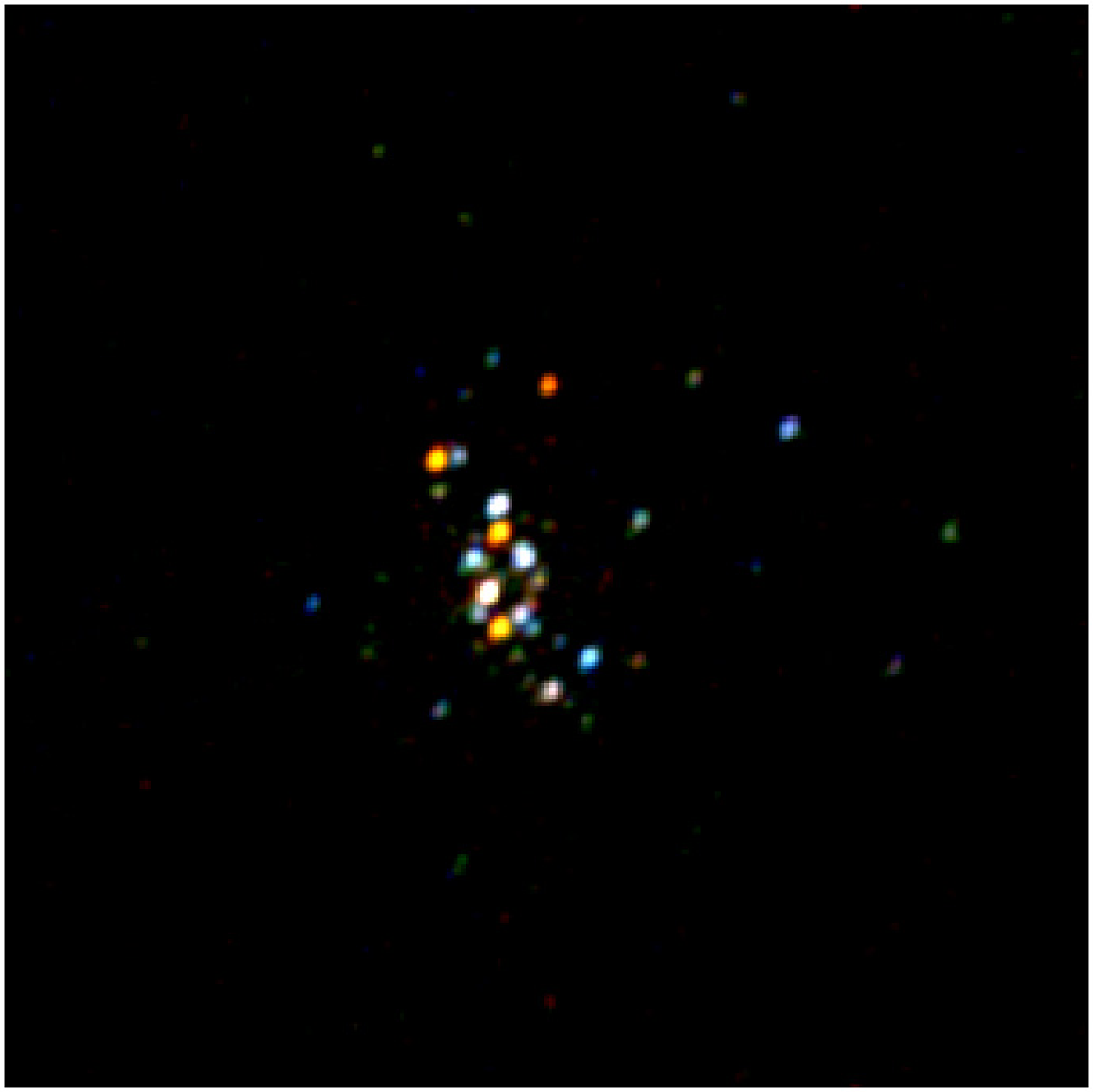}
\caption{\chandra\ image of the globular cluster NGC~6266.  This 63 ks
observation was made with the ACIS-S3 chip on 2002 May 12.  Photons in
the range 0.5--1.2~keV are shown in red, 1.2--2.5~keV photons in
green, and 2.5--6~keV photons in blue.  The image is 2\farcm46 on a
side, corresponding to the cluster's half-mass diameter.  Fifty-one
sources are detected in this observation using the wavelet-based
algorithm {\it wavdetect} supplied by the \chandra\ X-ray Center.  The
image has been smoothed by convolution with a two-dimensional Gaussian
with a full width at half maximum of 1\arcsec, which corresponds to
the telescope's point spread function. A more detailed analysis of
this data is in preparation.  \label{fig:ngc6266}}
\end{figure}

Because these systems are on average heavier than most other members
of a globular cluster, they sink in the cluster's gravitational
potential to the crowded centre, thus requiring high spatial
resolution telescopes to resolve them.  The advantage of \chandra's
sub-arcsecond spatial resolution is clearly seen in the image of the
globular cluster NGC~6266 (Fig.~\ref{fig:ngc6266}), one of the richest
clusters observed to date.  Fifty-one sources are detected within the
cluster's 1\farcm23 half-mass radius.  Approximately two or three are
background sources (see below). Not only can \chandra\ resolve the
sources, it also has the energy resolution to distinguish spectral
differences among them.  In the image, photons in the range 0.5--1.2
keV are shown in red, those in the range 1.2--2.5 keV are shown in
green, and those in the range 2.5--6 keV are in blue.

\begin{figure}
\centering
\plotone{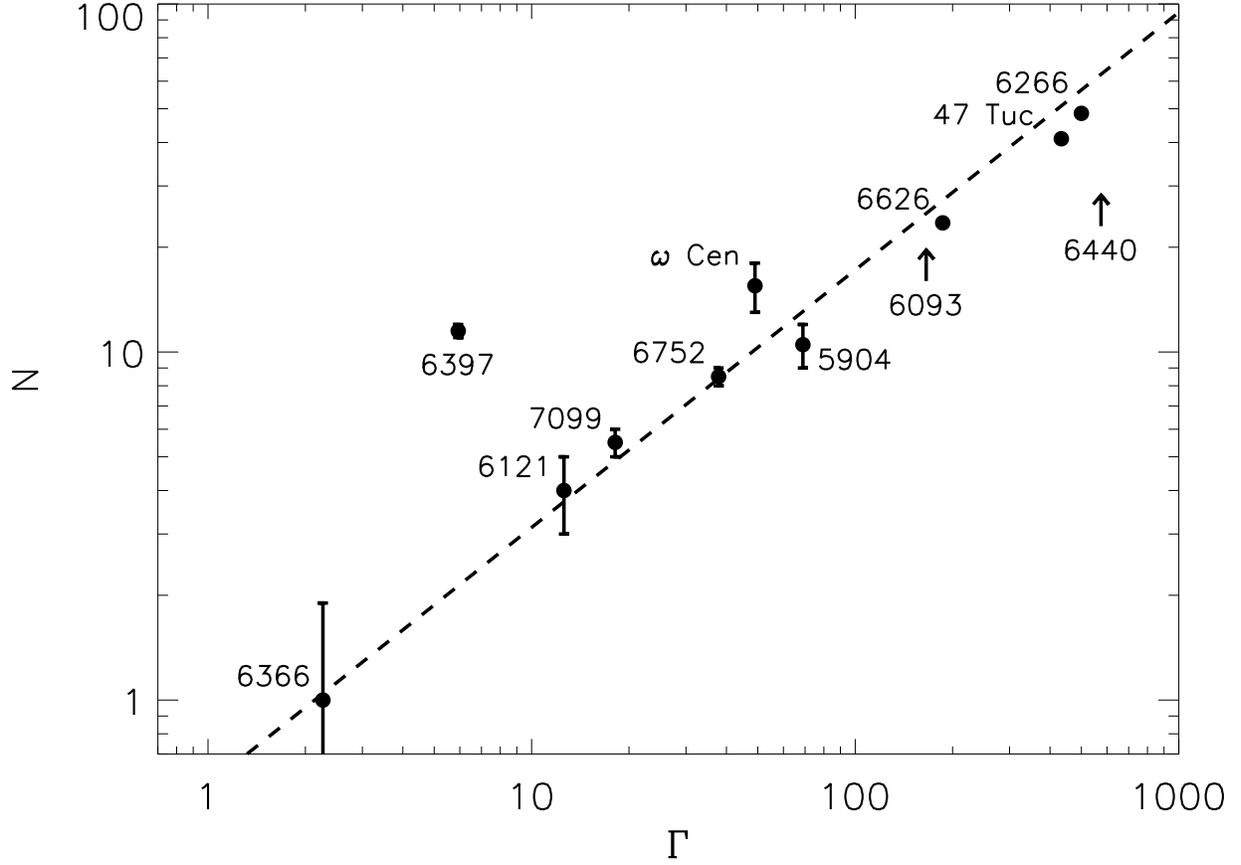}
\caption{Number of globular cluster X-ray sources ($N$) with $\Lx \ga
4\times10^{30}$~\ergsec\ versus the normalised encounter rate $\Gamma$
of the cluster.  The normalisation has been chosen such that
$\Gamma/100$ is roughly the number of LMXBs in a cluster or, for the
cases $\Gamma < 100$, the percent probability of the cluster hosting
an LMXB.  A {\boldmath $\uparrow$} indicates a globular cluster for
which the \chandra\ observation did not reach the required
sensitivity. \label{fig:nsrcs}}
\end{figure}

To explore the relationship between a cluster's physical properties
and its close binary population, we used published results and
available data for the 12 clusters so far observed with \chandra.  We
searched each cluster for sources to a limiting luminosity of about
$4\times10^{30}$~\ergsec\ (in the 0.5--6~keV range) using a
wavelet-based algorithm available from the \chandra\ X-ray Center.
(Two of the clusters --- NGC 6093 and NGC 6440 --- were not observed
long enough to reach this luminosity limit.)  To estimate the number
of sources associated with the cluster, we count all sources detected
within the half-mass radius of the cluster and subtract the estimated
number of background sources based on the $\log{N}-\log{S}$
relationship of \citet{gi01} The uncertainty in the number of sources
in a cluster in Fig.~\ref{fig:nsrcs} is due to the uncertainty in the
estimates for the number of background objects.

Following \citet{vh87}, we estimate the encounter rate per unit volume
($R$) of a cluster as $R\propto \frac{\rho^2}{v}$, where $\rho$ is
density and $v$ is velocity dispersion.  For each cluster, we perform
a volume integral of this quantity from the centre to the half-mass
radius to obtain our estimate for the encounter rate $\Gamma$.  The
forms of $\rho$ and $v$ as functions of radius are easily obtained
from the models developed by \citet{ki66}, which can be specified by
the parameters tabulated by \citet{ha96} in the 2003 February version
of his catalog.  The normalizations are set by the central values
$\rho_o$ and $v_o$ (this differs from the analysis of Verbunt \& Hut,
who estimated $v_o$ by the viral theorem), which we obtained from the
Harris catalogue and the catalogue of \citet{pm93}, respectively.

We have searched for correlations (using the Spearman $\rho$
correlation coefficient) between the number of X-ray sources in a
cluster ($N$) and physical parameters which we expect to be important
in determining $N$, such as encounter rate $\Gamma$, cluster mass $M$,
central density $\rho_o$, core radius $r_c$, and half-mass relaxation
time $t_h$.  In most cases, we find correlations, but the best is
between $N$ and $\Gamma$ (Table~\ref{tab:correl}).  The next best
correlation we find is with $M$.  Higher $M$ on average corresponds to
larger $N$, but most of that variation stems from the fact that
$\Gamma$ and $M$ are naturally correlated: keeping cluster size and
concentration parameter constant while increasing $M$ will increase
$\Gamma$.  If encounters were not to play a role in the formation of
X-ray sources, one would expect a tight (in first approximation
linear) correlation between $N$ and $M$ and a more loose correlation
between $N$ and $\Gamma$, contrary to what we find.  As an additional
check, we estimated $N$ for two clusters not observed deeply enough,
NGC 6093 and NGC 6440, by extrapolating their observed luminosity
functions, and we find that these estimates improve the correlation
with $\Gamma$ but worsen the correlation with $M$.


We plot $N$ versus $\Gamma$ (in normalized units described below) in
Fig.~\ref{fig:nsrcs}.  Clusters not observed deeply enough to reach
our luminosity limit are indicated by arrows.  Each point is
identified by the cluster's NGC designation or other name.  A power
law fit (not including the lower limits) indicates that
$N\propto\Gamma^{0.74}$, with errors on the power-law index of
$\pm$0.36.  This correlation offers the first quantitative, empirical
link between the encounter rate (over a range of almost three orders
of magnitude) and the number of exotic close binaries in a globular
cluster and suggests that most of these systems are formed dynamically
through some sort of encounter.


Note that there is one cluster, NGC 6397, that falls significantly
outside the otherwise good fit in Fig.~\ref{fig:nsrcs}.
Interestingly, this cluster has a high central density and a tidal
radius that is far smaller than would be expected for its current
location relative to the Galactic centre, suggesting that it describes
a highly eccentric orbit around the Galactic centre, which has caused
large tidal stripping during each perigalacticon passage.  The
combination of a high central density as a good place to make X-ray
sources, together with significant tidal mass loss, would create an
efficient distillation process whereby the binaries remain in the core
and many single stars are stripped from the outer regions of the
cluster, leading to a `high grade' cluster enriched in X-ray sources,
which is exactly what is observed.

The relationship in Fig.~\ref{fig:nsrcs} deals with a mixture of (at
least) four different kinds of sources (quiescent LMXBs, CVs, MSPs,
main sequence binaries), all of which are expected theoretically to be
primarily formed through encounters in globular clusters.  These
expectations are now confirmed by the evidence presented in
Fig.~\ref{fig:nsrcs} and Table~\ref{tab:correl}.  Note that there are
many remaining uncertainties concerning precise theoretical
predictions of formation rates of LMXBs, CVs, and MSPs.  For each
separate category there are several different formation channels, such
as tidal capture, exchange reactions involving encounters between
single stars and binaries, or between binaries and binaries.  The only
good way to get a quantitative handle on the whole mix is to do
detailed simulations for individual clusters \citep{b03a,b03b}.

Bypassing these complexities, the simple encounter frequency adopted
here, density squared divided by velocity, describes how often a
cluster member comes close to another, taking into account
gravitational focusing.  First of all, it does not discriminate
between different objects (MS stars, giants, WDs, NSs, and binaries of
all types); and secondly it neglects possible velocity dependencies in
3-body and 4-body interactions.  If there were no correlations between
the abundances of objects involved in encounters with, say, total mass
of a cluster, then we would expect encounters between two single stars
to be proportional to $\Gamma$, hence $N$ would be linearly
proportional to $\Gamma$.  The result $N\propto\Gamma^{0.74\pm0.36}$
is consistent with the simplest prediction, a slope of unity.

The next important step is to examine this relationship for each
individual class of objects, but this requires identifying each of the
$\sim$200 sources represented in Fig.~\ref{fig:nsrcs}, which is an
ongoing and very time-consuming process, for which the X-ray data
alone are not sufficient.  Only three clusters so far --- 47 Tuc
\citep{gr01a}, NGC 6397 \citep{gr01b}, and NGC 6752 \citep{po02a} ---
have had the X-ray (\chandra), optical ({\it Hubble Space Telescope}),
and radio data necessary to identify a substantial number of sources.
About 50\% of the sources in 47~Tuc, 75\% of those in NGC~6397, and
80\% of those in NGC 6752 have been identified to date.

However, it has become clear that \chandra\ data alone are sufficient
to identify the quiescent LMXBs in a cluster as distinct from the
other three source types based on their luminosities and broad-band
spectral properties.  In a globular cluster, only LMXBs and CVs are
more luminous than $10^{32}$~\ergsec, and quiescent LMXBs have a much
softer spectrum than CVs so that a ratio of the number of photons
detected in a soft band (0.5--1.5 keV) to the number detected in a
hard band (1.5--6 keV) suffices to distinguish the two.  We have
tested these selection criteria on the securely identified quiescent
LMXB in NGC 6440 \citep{po02b} as well as known quiescent LMXBs not
located in globular clusters: Aql X-1, Cen X-4, MXB 1659$-$298, KS
1731$-$260, and 4U 2129$+$47.  Using archival \chandra\ data, we find
that the criteria successfully identify all of them.

We apply these selection criteria to the twelve clusters observed with
\chandra\ and use the results of \xmm\ observations of NGC 6205
\citep{ge03} and NGC 6656 \citep{we02} to determine the LMXB content
of 14 globular clusters.  A total of 19--22 LMXBs have been found in
these clusters, with some having multiple LMXBs and some having none.
A picture is emerging which appears to confirm the idea of
\citet{vh87} that the number of LMXBs is proportional to the encounter
frequency ($\Gamma$) of the cluster.  A power-law fit similar to that
in Fig.~\ref{fig:nsrcs} was done for the globular clusters containing
{\it multiple} LMXBs, and we find that the best-fit power-law index is
0.97, indicating a nearly linear relationship.  The errors on the
power-law index are rather large ($\pm$0.5) because of the small
number of LMXBs (15) involved in the fit.  A similar correlation was
reported by \citet{ge03}, who assumed a linear relationship {\it a
priori}.  The (nonparametric) Spearman $\rho$ correlation coefficient
between the number of LMXBs and $\Gamma$ is 1.0, and the Pearson $r$
linear correlation coefficient is 0.85, indicating a high degree of
linear correlation.  We take the relationship to be linear for the
following discussion.

We have normalized $\Gamma$ in Fig.~\ref{fig:nsrcs} such that
$\Gamma/100$ is roughly the number of LMXBs in the cluster.  The
interpretation of $\Gamma$ in clusters with low encounter rate is then
the percent probability that the cluster will host an LMXB.  For
example, for every cluster like NGC~7099 (with $\Gamma\approx20$)
which hosts an LMXB, there should be, on average, four similar
clusters which do not.  It is therefore not surprising to see a few
LMXBs in clusters with low encounter frequencies.

To estimate the total number of LMXBs expected in the 140 known
Galactic globular clusters, we simply need to add the encounter rates
of all clusters.  However, our method for estimating $\Gamma$ is
applicable to only about one-third of the clusters since $v_o$ is
known for only that many.  We therefore use the estimate for $\Gamma$
described by \citet{vb02}, in which the volume integral of $R$ is
taken only out to the core radius, over which the density and velocity
dispersion are roughly constant.  Then, $\Gamma \propto
\frac{{\rho_o}^2{r_c}^3}{v_o}$.  The virial theorem relates $v_o$,
$\rho_o$, and $r_c$ via $v_o \propto \sqrt{\rho_o}r_c$.  Therefore, we
can estimate the encounter rate for each cluster by $\Gamma \propto
{\rho_o}^{1.5}{r_c}^2$.

We again use the catalog of \citet{ha96} for these parameters, with
the updates for Terzan~5 by \citet{he03} From adding the encounter
rates of all clusters ($\Gamma_{\rm tot}$), we estimate that roughly
100 LMXBs reside in our Galaxy's globular clusters.  Of these 100
expected LMXBs, there are 13 persistently or transiently bright
globular cluster LMXBs, most of which have been known for almost 20
years.  With the 19--22 quiescent LMXBs discovered by \chandra\ and
\xmm\ in the past few years, the known population is about one-third
of the expected total.  We can check for consistency in the following
way.  The encounter rates of the four clusters whose 15 LMXBs were
used in the power-law fit mentioned above add up to 15\% of
$\Gamma_{\rm tot}$, by definition.  The summed encounter rate of the
other 10 clusters which have been observed deeply enough to determine
their entire LMXB content is 5\% of $\Gamma_{\rm tot}$.  These
clusters host between 4 and 7 LMXBs, in good agreement with our
predictions.  Efforts are underway to uncover the rest of the expected
LMXB population in globular clusters.  These numbers will prove
extremely useful in testing models of cluster evolution and LMXB
formation.

\acknowledgments
DP and WHGL gratefully acknowledge support from NASA.

\clearpage

\clearpage

\clearpage

\begin{deluxetable}{ccc}
\tablewidth{0pt}
\tablecaption{Spearman $\rho$ correlation coefficients of $N$ versus
various cluster properties. \label{tab:correl}}
\tablecolumns{3}
\tablehead{
\colhead{Parameter} &\colhead{Spearman $\rho$} & \colhead{Probability\tablenotemark{1}}
}
\startdata
$\Gamma$&	0.855&	0.9984\\
$M$&		0.758&	0.9889\\
$t_h$&        	0.588&      0.9261\\                                              
$\rho_o$&      	0.418&      0.7709\\                                              
$r_c$&       	$-$0.054&     0.1190\\ 
\enddata
\tablenotetext{1}{Probability that Spearman $\rho$ is different than
zero.  A correlation coefficient of zero corresponds to the data being
uncorrelated.}
\end{deluxetable}

\end{document}